\documentclass{iau}
\usepackage{natbib}
\usepackage{amsmath}
\usepackage{graphicx}
\usepackage{multirow}
\usepackage{bm}
\newcommand\apj{{Astrophys. J}}%
\newcommand\apjl{{Astrophys. J Lett.}}%
\newcommand\ssr{{Space~Sci.~Rev.}}%
\newcommand\mnras{{MNRAS}}%

\def\Rs{R_{s}}
\newcommand{\bl}{Babcock--Leighton}

\newcommand{\Fig}[1]{Fig.~\ref{#1}}
\newcommand{\Eq}[1]{Eq.~(\ref{#1})}

\def\Rs{R_{s}}
\newcommand{\etasurf}{\eta_{\mathrm{surf}}}
\newcommand{\etaSCZ}{\eta_{\mathrm{SCZ}}}
\newcommand{\etaRZ}{\eta_{\mathrm{RZ}}}
\newcommand{\rBCZ}{r_{\mathrm{BCZ}}}
\newcommand{\rsurf}{r_{\mathrm{surf}}}

\begin{document}

\lefttitle{V. Vashishth}
\righttitle{Cycle variability and grand minima}

\jnlPage{1}{7}
\jnlDoiYr{2024}
\doival{10.1017/xxxxx}

\aopheadtitle{Proceedings IAU Symposium No. 365}
\editors{A. V. Getling, \&  L. L. Kitchatinov, eds.}

%\title{Modelling the long-term variability of sun-like stars: From subcritical to supercritical dynamos}
\title{Modelling the rotation dependence 

of cycle variability in sun-like stars: 
Answering why only slowly rotating stars produce grand minima}
%\title{Modelling the rotation dependence of the frequency of grand minima and cycle variability of sun-like stars: Answering why only slowly rotating stars produce grand minima}

\author{Vindya Vashishth$^1$}
\affiliation{$^1$Department of Physics, Indian Institute of Technology (Banaras Hindu University) Varanasi 221005 India \\ email: {\tt vindyavashishth.rs.phy19@itbhu.ac.in}}

\begin{abstract}
The Sun and 
solar-type 
stars exhibit irregular cyclic variations in their magnetic activity over long time scales. 
To understand this irregularity, we employed the flux transport dynamo models to investigate the behavior of one solar mass star at various rotation rates. To achieve this, we have utilized a mean-field hydrodynamic model to specify differential rotation and meridional circulation, and we have incorporated stochastic fluctuations in the Babcock-Leighton source of the poloidal field to capture inherent fluctuations in the stellar convection. Our simulations successfully demonstrated consistency with the observational data, revealing that rapidly rotating stars exhibit highly irregular cycles with strong magnetic fields and no Maunder-like grand minima. On the other hand, slow rotators produce smoother cycles with weaker magnetic fields, long-term amplitude modulation, and occasional extended grand minima. We observed that the frequency and duration of grand minima increase with the decreasing rotation rate. These results can be understood as the tendency of a less supercritical dynamo in slow rotators to be more prone to produce extended grand minima.
We further explore the possible existence of the dynamo in the subcritical regime in a Babcock-Leighton-type framework and in the presence of a small-scale dynamo.
\end{abstract}


\maketitle
\vspace{-0.5cm}
\section{Introduction}

Like our Sun, many other sun-like stars have magnetic fields and cycles as unveiled by various observations \citep{donati92, Baliu95, wright11, WD16, Travis16}.
According to these observations, a star's rotation rate plays an important role in determining its magnetic activity. Rapidly rotating (young) Sun-like stars exhibit a high activity level with no Maunder-like grand minimum (flat activity) and rarely display smooth regular activity cycles. On the other hand, slowly rotating old stars like the Sun and older have lower activity levels and smooth cycles with occasional grand minima \citep{Skumanich72, R84, Baliu95, Olah16, BoroSaikia18, garg19}.
Recently, \citet{Shah18} observed the decreasing magnetic activity of HD 4915, which might indicate it as the Maunder minimum candidate. Later \citet{anna2022} confirmed that HD 166620 is entering into a grand minimum phase. Interestingly, these stars (including Sun) are slow rotators.

Magnetic cycles in the Sun and other sun-like stars are maintained by dynamo action powered by helical convection and differential rotation in their convection zones \citep{Kar14a, Cha20, Kar23}. 
This is because the toroidal field is generated through the stretching of the poloidal field by the differential rotation, which is known as the  $\Omega$ effect. 
There is strong evidence that the poloidal field is generated through a mechanism, so-called the Babcock--Leighton process \citep{Das10, KO11, Muno13, Priy14, Mord20, Mord22, KKV21}. 
In this process, tilted sunspots \citep[more generally bipolar magnetic regions;][]{Anu23} decay and disperse to produce a poloidal field through turbulent diffusion, meridional flow, and differential rotation.
While the systematic tilt in the BMR is crucial to generate the poloidal field, the scatter around Joy's law tilt \citep[e.g,][]{MNL14, Jha20} produces a variation in the solar cycle \citep{LC17, KM17, KM18, KMB18, Kar20, BKK23}.   

Many observations suggest that as the stars rotate faster, the magnetic activity becomes stronger, but the relation between the activity cycle period and the rotation rate does not seem a straightforward trend. The cycle period tends to decrease with the rotation rate for the slowly rotating stars, whereas the trend is quite complicated for the fast rotators. Previous studies have explored the trend of magnetic field strength and the cycle period with the rotation rate of the stars \citep{KKC14, Hazra19}. \citet{KTV20, nora22} have also explored the possibility of magnetic cycle and reversals in slowly rotating stars possibly having anti-solar differential rotations \citep[e.g.,][]{Kar15, KMB18}. Here we aim to understand these observational trends of stellar magnetic activity using dynamo modeling.
We shall extract the dependency of the rotation rate of the sun-like stars on its cycle variability and the occurrence of the grand minima using the dynamo models of \cite{KKC14, Hazra19} in which the regular behavior of the stellar cycle was simulated. As the stellar cycles are irregular, it is natural to explore the irregular features of the stellar cycles using these models.
For this, we have included stochastic noise to capture the inherent fluctuations in the stellar convection, as seen in the form of variations in the flux emergence rates and the tilts of BMRs around Joy’s law. To do so, we have included the stochastic fluctuations in the \bl\ source for the poloidal field in the dynamo.

\section{Model}

In our work, we have developed three kinematic mean-field dynamo models, namely, Models I-III, by assuming the axisymmetry. Thus, the evolution equation of the poloidal ($\nabla \times [ A(r, \theta) {\bf e}_{\phi}]$) and toroidal ($B (r, \theta) {\bf e}_{\phi}$) fields are followings.

\begin{equation}
\frac{\partial A}{\partial t} + \frac{1}{s}({\bm v_p}\cdot \bm \nabla)(s A)
= \eta \left( \nabla^2 - \frac{1}{s^2} \right) A + S(r,\theta;B),
\label{eq:pol}
\end{equation}

\vspace{-0.2in}

\begin{equation}
\frac{\partial B}{\partial t}
+ \frac{1}{r} \left[ \frac{\partial}{\partial r}
(r v_r B) + \frac{\partial}{\partial \theta}(v_{\theta} B) \right]
= \eta \left( \nabla^2 - \frac{1}{s^2} \right)B \nonumber
+ s(\bm B_p \cdot \bm \nabla) \Omega +
\frac{1}{r}\frac{d\eta}{dr}\frac{\partial{(rB)}}{\partial{r}},
\label{eq:tor}
\end{equation}
where $s = r \sin \theta$, ${\bm v_p} = v_r {\bm \hat{ r}} + v_\theta {\bm \hat{ \theta}}$ is the meridional flow and the $\Omega$ is the angular velocity whose profile is obtained from the mean-field hydrodynamic model of \cite{KO11}, $\eta$ is the turbulent magnetic diffusivity which is written as the function of $r$ alone and take the following form,
\begin{eqnarray}
\eta(r) = \etaRZ + \frac{\etaSCZ}{2}\left[1 + \mathrm{erf} \left(\frac{r - \rBCZ}
{d_t}\right) \right]
+\frac{\etasurf}{2}\left[1 + \mathrm{erf} \left(\frac{r - \rsurf}
{d_2}\right) \right]
\label{eq:eta}
\end{eqnarray}\\
with $\rBCZ=0.7 \Rs$ ($\Rs$ being the stellar radius), $d_t=0.015 \Rs$, $d_2=0.05 \Rs$, $\rsurf = 0.95 \Rs$,
$\etaRZ = 5 \times 10^8$ cm$^2$~s$^{-1}$, $\etaSCZ = 5 \times 10^{10}$ cm$^2$~s$^{-1}$, and
$\etasurf = 2\times10^{12}$ cm$^2$ s$^{-1}$.$S$ is the source for the 
poloidal field and its parameterized form is written as 

\begin{equation}
 S(r, \theta; B) = \frac{\alpha_0 \alpha_{\rm BL}(r,\theta)}{1 + \left( \overline{B} (r_t,\theta)/B_0 \right)^2} \overline{B}(r_t,\theta),
\label{source}
\end{equation}

where $\overline{B}(r_t,\theta)$ is the toroidal field at latitude $\theta$
averaged over the whole tachocline from $r = 0.685 \Rs$ to $r=0.715 \Rs$, 
$\alpha_0$ is the measure of the strength of the \bl\ process, which is expressed as the dependence on the rotation in the following way,
\begin{eqnarray}
 \alpha_0 = \alpha_{0,s} \frac{T_s}{T},
 \label{eq:alpha0I}
\end{eqnarray}
where $\alpha_{0,s}$ is the value of $\alpha_0$
for the solar case, which is taken as $0.7$ cm~s$^{-1}$ in Model I-II and $T_s$ and $T$ are the rotation period of Sun and the star, respectively.
And finally,  
$\alpha_{\rm BL}$ is the parameter for Babcock--Leighton process which is written in Model I-II as,
\begin{equation}
\alpha_{\rm BL}(r,\theta)=\frac{1}{4}\left[1+\mathrm{erf}\left(\frac{r-r_4}{d_4}\right)\right]\left[1-\mathrm{erf}\left(\frac{r-r_5}{d_5}\right)\right]\times \sin\theta\cos\theta
\label{alpha}
\end{equation}
where, $r_4=0.95 \Rs$, $r_5= \Rs$, $d_4=0.05 \Rs$, $d_5=0.01 \Rs$, and for Model III, the $\alpha$ profile used is given by,
\begin{eqnarray}
\alpha_{\rm BL}(r,\theta)=\frac{1}{2}\left[1+\mathrm{erf}\left(\frac{r-\rsurf}{d}\right)\right] \sin ^{2}\theta\cos\theta,
\label{alpha_model3}
\end{eqnarray}
where $d= 0.01\Rs$. 

Thereafter, to study large-scale magnetic fields, we included fluctuations in the source of the poloidal field. The main reason for including the randomness was that since the star cycle's amplitude is not equal, it varies from time to time. This happens due to the fluctuating nature of the stellar convection. The dynamo parameters fluctuate around their mean.
In \bl, the fluctuations are due to the scatter in the bipolar active region tilts around the Joy's law. This randomness changes the poloidal field and makes irregular magnetic cycles as observed in Sun and Sun-like stars. In order to include randomness in our \bl\ $\alpha$, we include fluctuations in the $\alpha$ appearing in \Eq{eq:alpha0I} as, $\alpha_{0,s} = \alpha_{0,s} r, $  
where $r$ is the Gaussian random number with mean unity and standard deviation ($\sigma$) as 2.67. We keep the value of $\sigma$ the same for all the stars.
In our models, the value of $\alpha_0$ is updated randomly after a certain time, which we take to be one month.

The $\alpha_0$ in all three models have the same form  (\Eq{eq:alpha0I}), except in Model III, $\alpha_{0,s} = 4$ cm~s$^{-1}$ and fluctuations in this model are included separately in the two hemispheres. We note that above $\alpha$ in \Eq{alpha_model3} has a $\sin ^{2}\theta\cos\theta$ dependence instead of $ \sin\theta\cos\theta$ as used in Models I-II to make the $\alpha$ effect strong (weaker) in low (high) latitudes. Also, the radial 
extent of this $\alpha$ is a bit wider than that used in Models I-II. 

Finally, In Model III, we have added radial magnetic pumping. 
This inclusion of pumping is inspired by \citet{Hazra19}, who found some agreement of the cycle period vs rotation trend with observations. 
It was realized that a downward magnetic pumping helps to make the magnetic field radial near the surface and reduce the toroidal flux loss through the surface, making the dynamo model in accordance with the surface flux transport models and observations \citep{Ca12, KC16}. 
%The near-surface pumping also helps the dynamo to operate at a high diffusivity range consistent with the mixing-length theory \citep{KO12, KC16, KM17}, and facilitates the model to recover from the Maunder-like extended grand minima \citep{KM18}. 
The pumping has the following form:
\begin{equation}
\gamma = - \gamma_0 \left[1 + \mathrm{erf} \left(\frac{r - 0.9\Rs}
{0.02\Rs}\right) \right],
\label{pumping}
\end{equation}
where the amplitude of the radial magnetic pumping is given by $\gamma_0$ which is 24 m\,s$^{-1}$ in all the stars.

\section{Results $\&$ Discussion}
\begin{figure}
% \vspace*{-2.0 cm}
%\begin{center}
\begin{minipage}[t]{0.65\textwidth}
\centering
\includegraphics[scale= 0.55]{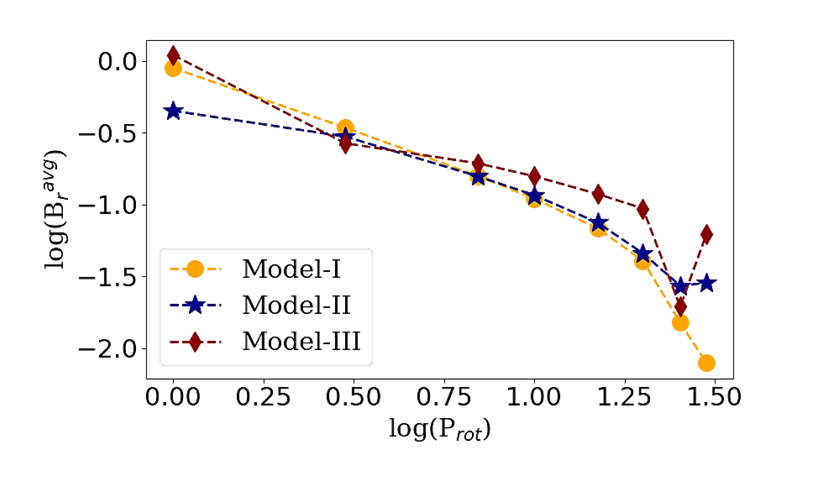} 
% \vspace*{-1.0 cm}
  \end{minipage}%
  \begin{minipage}[t]{0.35\textwidth}
  \vspace{-4.2cm}
 \caption{Variation of the magnetic activity with the rotation period of stars for all three models. Figure adopted from \citet{Vindya23}.}
   \end{minipage}
   \label{fig1}
%\end{center}
\end{figure}

The simulations were done for $M_\odot$ mass stars having rotation periods of 1, 3, 7, 10, 15, 20, 25.30 (Sun), and 30 days, respectively.
Here we discuss the various aspects of magnetic cycles obtained from all the stars and from all the three models.
%Firstly, in Model I-II, we find the regular polarity reversal from the time-latitude plot of the toroidal field at the base of the convection zone for all the rotation periods in all the models as shown in Fig.\,\ref{fig1}. This figure also depicts a weak equatorward propagation of the field at lower latitudes which is due to the transport of the field by the equatorward meridional circulation.

Firstly, the regular polarity reversals in the toroidal field were noted. Model II presents an irregular magnetic cycle with significant hemispheric asymmetry. In contrast, Model III shows regular polarity reversals for fast rotators and the Sun but irregularity for slower ones. Depending on the rotation period, stars exhibit varying magnetic field configurations: slow rotators mainly have dipolar fields, while those rotating in 7 days or less can have quadrupolar configurations. The pictorial representation and detailed analysis are in \citet{Vindya23}.
%Additionally, slowly-rotating stars display an equatorward migration of the toroidal field, while fast rotators have a weak poleward shift. Notably, fast-rotating stars possess stronger magnetic fields due to an increased rotation rate. There's a noted discrepancy between the model's predictions and observational data on field strength in rapidly rotating stars. Lastly, slowly-rotating stars display long-term variations in their magnetic cycles, while fast rotators adjust quickly after a weak phase. The findings are supported by various referenced researches and visualized in three time-latitude plots.

One obvious feature in these simulations, as seen in \Fig{fig1}, is that the magnetic field becomes strong in fast-rotating stars. 
This is because the strength of $\alpha$ increases with the rotation rate of the star (the shear, however, remains more or less unchanged in different stars). 
If a star rotates faster, the tilt of the BMR associated with the \bl\ $\alpha$ is expected to increase. Therefore, with age, as the rotation rate decreases, the dynamo process becomes weaker, and the dynamo number also decreases. This implies that the rapidly rotating stars will likely have stronger magnetic cycles.
This result agrees with the \cite{KKC14} and the observations \citep{Noyes84a, wright11}. 
%We also observe that slowly rotating stars are more irregular as compared to the fast rotating stars.

%One can also hint at the persistence from the time-latitude plots observed in different stars. 
%Slowly rotating stars seem to produce more long-term modulation in their cycles, including extended episodes of a weaker magnetic field. 
%In contrast, fast rotators generate less long-term modulation.

%The root cause for such behavior is that the slowly rotating stars have a small dynamo number. Due to this, if the magnetic cycle gets weaker sometimes, it would take a long time to grow the field. Therefore, we see a long-term modulation in the slowly rotating stars. But for the fast rotators, the cycle gets stronger much more quickly after getting into the weak phase. This is easy to understand, as, in fast rotators, the dynamo number is high so the growth rate is very high. This trend is also explained in \cite{KKV21, Vindya21}. The results are in accordance with the observations as well \citep{Baliu95}. 

\begin{figure}
% \vspace*{-2.0 cm}
%\begin{center}
\centering
\includegraphics[scale= 0.7]{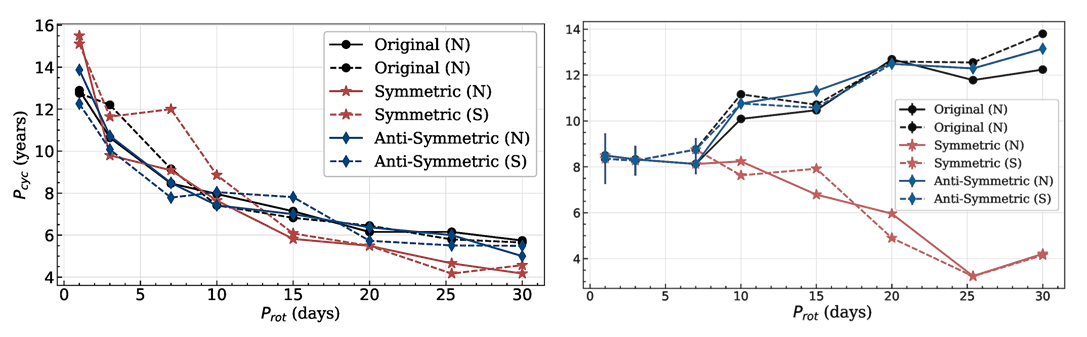} 
% \vspace*{-1.0 cm}
 \caption{Change in the stellar cycle duration with the rotation period of stars for (a) Models I-II and (b) Model III. From \citet{Vindya23}.}
   \label{fig2}
%\end{center}
\end{figure}

We also computed the cycle periods for all three models. To achieve this, we analyzed the Fourier power spectrum peaks of the toroidal field time series within the tachocline. This analysis was conducted separately for the northern and southern hemispheres, distinguishing between symmetric and anti-symmetric cycles.
The variations of the cycle duration in each case with the rotation rate are shown in \Fig{fig2}.
Notably, Models I and II exhibit an increasing trend in the cycle period with higher stellar rotation rates. This behavior is due to the weakening of the meridional flow with the decrease in the rotation period which leads to intensifying the flow speed in the thin layers near the boundaries.
%This happens because, as the rotation period decreases (or rotation rate increases), the meridional flow becomes weaker (although the flow speed increases in the thin layers near the top and bottom boundaries). 

Although these two models reproduced various stellar observations, they failed to reproduce the magnetic cycle period vs. rotation trend correctly for the slowly-rotating stars. One way to resolve this discrepancy was to include 
radial magnetic pumping in the stellar CZs as done by \citet{Hazra19}.
As a result, in Model III, we got the cycle-rotation period trend closer to the observations. 

When strong downward magnetic pumping is included in this model, the diffusion of the magnetic field across the surface becomes negligible, and then the dynamo allows it to operate at a low $\alpha$ \citep{KC16}. The lower the $\alpha$, the longer the cycle period. We can see from \Fig{fig2} that at 30 days rotation period, while Models I-II were producing a cycle period of 6 years, Model III produced a much longer period of 13 years. Then with the decrease of the rotation period, the $\alpha$ becomes stronger and thus the poloidal field generation process becomes more efficient. This makes the reversal of the field faster. This effect in the pumping-dominated regime overpowers the increase of the cycle period due to a decrease in meridional flow speed.

\begin{figure}
% \vspace*{-2.0 cm}
%\begin{center}
\centering
\includegraphics[scale= 0.38]{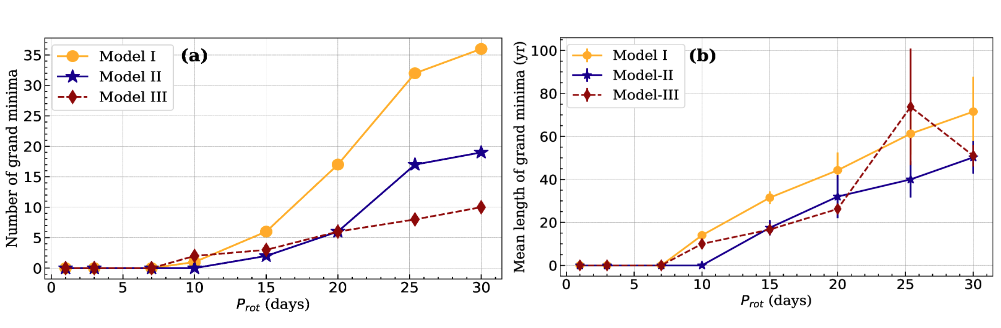} 
% \vspace*{-1.0 cm}
 \caption{Change of (a) the number and (b) the average duration of grand minima with the rotation period of stars. Yellow circles, blue asterisks, and red diamonds depict Models I, II, and III trends, respectively. In (b), the error bars are computed from the standard deviation of the durations of the grand minima in each case. From \citet{Vindya23}.}
   \label{fig3}
%\end{center}
\end{figure}

The focal point of our study was to understand how the long-term variability of stellar cycles is influenced by their rotation rates. We found that fast rotators manifest irregular cycles with less pronounced long-term variability, while slower rotators exhibit longer modulations in their cycles, interspersed with periods of weaker magnetic fields.
We then identified extended periods of low magnetic activity, known as grand minima, using a method adapted from solar studies by \citet{USK07}. We examined the number of grand minima observed in each case with the help of a time-series plot of the toroidal field at the base of CZ and radial magnetic field from simulations for 11,000 years. We infer that the number of grand minima observed in all the models shows an increasing trend with the rotation period. We saw that the rapidly rotating stars hardly produce any grand minima, whereas the slowly rotating stars produce some grand minima, and also, as the rotation period increases, the number of grand minima is seen to increase (see Fig.\,\ref{fig3}). This is because, with the increase of rotation period, the supercriticality of the dynamo decreases, and the dynamo is more prone to produce extended grand minima in this regime. 
This result is as per \cite{Vindya21} where we observed that as the supercriticality increases (i.e., as the dynamo number increases), the frequency of occurrence of grand minima decreases.
This is also supported by the observation that the detected grand/Maunder minima candidates are the slow rotators.
Notably, the average duration of these grand minima also increases with rotation duration, with most lasting under 150 years and the majority even below 70 years.

\vspace{-0.2in}
\begin{flushleft}
{\bf \item[4.]Conclusion}
\end{flushleft}
\vspace{-0.1in}
Based on the kinematic dynamo simulations of one solar mass at different rotation rates with stochastically forced \bl\ source, we make the following conclusions.
In slowly rotating stars, the cycles are smooth and show long-term variation with occasional grand minima. Whereas the magnetic field is strong for rapidly rotating stars, cycles are more irregular, and no grand minima are detected. The number of grand minima increases with the decrease in the star's rotation rate. Details of this work have been presented in \citet{Vindya23}. We further explore the possible existence of the dynamo in the subcritical regime in a Babcock-Leighton-type framework and in the presence of a small-scale dynamo, whose details have been presented in \cite{Vindya21} and Vashishth et al. 2023 (under preparation).

%\bibliographystyle{iaulike}
%\bibliography{iauguide}
%\bibliography{paper} % if your bibtex file is called exampl

\end{document}